\documentclass[12pt]{article}
\newcommand{\nc}{\newcommand}
\nc{\renc}{\renewcommand}

\usepackage{calc}
\usepackage{ifthen}
\usepackage{graphicx}
\usepackage{epsfig}
\usepackage{miniplot}

\usepackage{wrapfig}
\usepackage{subfigure}
\usepackage{rotfloat}

\usepackage{epsfig}
\usepackage{graphicx}
\usepackage{epsf}

\nc{\half}{{\textstyle{1\over2}}}
\nc{\etal}{\mbox{\it et al. }}
\nc{\ie}{{\it i.e.}}
\nc{\eg}{{\it e.g.}}

\renc{\thefootnote}{\arabic{footnote}}
\nc{\capt}[1]{{\bf Figure.} {\small\sl #1}}

\nc{\eqs}[2]{\mbox{Eqs.~(\ref{#1},\,\ref{#2})}}
\nc{\eq}[1]{\mbox{Eq.~(\ref{#1})}}

\nc{\figs}[2]{\mbox{Figs.~(\ref{#1},\,\ref{#2})}}
\nc{\fig}[1]{\mbox{Fig~.(\ref{#1})}}

\nc{\tag}[1]{\label{#1} \marginpar{{\footnotesize #1}}}
\nc{\mtag}[1]{\label{#1} \mbox{\marginpar{{\footnotesize #1}}}}
\renc{\baselinestretch}{1.5}
\newlength{\overeqskip}
\newlength{\undereqskip}
\nc{\be}[1]{\begin{equation} \mbox{$\label{#1}$}}
\nc{\bea}[1]{\begin{eqnarray} \mbox{$\label{#1}$}}
\nc{\Section}[2]{\section{#2}\label{#1}}
\nc{\Bibitem}[1]{\bibitem{#1}}
\nc{\Label}[1]{\label{#1}}

\nc{\eea}{\vspace{\undereqskip}\end{eqnarray}}
\nc{\ee}{\vspace{\undereqskip}\end{equation}}
\nc{\bdm}{\begin{displaymath}}
\nc{\edm}{\end{displaymath}}
\nc{\dpsty}{\displaystyle}
\nc{\bc}{\begin{center}}
\nc{\ec}{\end{center}}
\nc{\ba}{\begin{array}}
\nc{\ea}{\end{array}}
\nc{\bab}{\begin{abstract}}
\nc{\eab}{\end{abstract}}
\nc{\btab}{\begin{tabular}}
\nc{\etab}{\end{tabular}}
\nc{\bit}{\begin{itemize}}
\nc{\eit}{\end{itemize}}
\nc{\ben}{\begin{enumerate}}
\nc{\een}{\end{enumerate}}
\nc{\bfig}{\begin{figure}}
\nc{\efig}{\end{figure}}
\nc{\arreq}{&\!=\!&}
\nc{\arrmi}{&\!-\!&}
\nc{\arrpl}{&\!+\!&}
\nc{\arrap}{&\!\!\!\approx\!\!\!&}
\nc{\non}{\nonumber\\*}
\nc{\align}{\!\!\!\!\!\!\!\!&&}

\def\lsim{\; \raise0.3ex\hbox{$<$\kern-0.75em
      \raise-1.1ex\hbox{$\sim$}}\; }
\def\gsim{\; \raise0.3ex\hbox{$>$\kern-0.75em
      \raise-1.1ex\hbox{$\sim$}}\; }
\nc{\DOT}{\hspace{-0.08in}{\bf .}\hspace{0.1in}}
\nc{\Laada}{\hbox {$\sqcap$ \kern -1em $\sqcup$}}
\nc\loota{{\scriptstyle\sqcap\kern-0.55em\hbox{$\scriptstyle\sqcup$}}}
\nc\Loota{{\sqcap\kern-0.65em\hbox{$\sqcup$}}}
\nc\laada{\Loota}
\nc{\qed}{\hskip 3em \hbox{\BOX} \vskip 2ex}

\nc{\real}{{\rm I \! R}}
\nc{\Z}{{\sf Z \!\!\! Z}}
\nc{\complex}{{\rm C\!\!\! {\sf I}\,\,}}
\def\bigid{\leavevmode\hbox{\small1\kern-3.8pt\normalsize1}}
\def\id{\leavevmode\hbox{\small1\kern-3.3pt\normalsize1}}
\nc{\slask}{\!\!\!/}
\nc{\bis}{{\prime\prime}}
\nc{\pa}{\partial}
\nc{\na}{\nabla}
\nc{\ra}{\rangle}
\nc{\la}{\langle}
\nc{\goto}{\rightarrow}
\nc{\swap}{\leftrightarrow}

\nc{\EE}[1]{ \mbox{$\cdot10^{#1}$} }
\nc{\abs}[1]{\left|#1\right|}
\nc{\at}[2]{\left.#1\right|_{#2}}
\nc{\norm}[1]{\|#1\|}
\nc{\abscut}[2]{\Abs{#1}_{\scriptscriptstyle#2}}
\nc{\vek}[1]{{\rm\bf #1}}
\nc{\integral}[2]{\int\limits_{#1}^{#2}}
\nc{\inv}[1]{\frac{1}{#1}}
\nc{\dd}[2]{{{\partial #1}\over{\partial #2}}}
\nc{\ddd}[2]{{{{\partial}^2 #1}\over{\partial {#2}^2}}}
\nc{\dddd}[3]{{{{\partial}^2 #1}\over
        {\partial #2 \partial #3}}}
\nc{\dder}[2]{{{d #1}\over{d #2}}}
\nc{\ddder}[2]{{{d^2 #1}\over{d {#2}^2}}}
\nc{\dddder}[3]{{d^2 #1}\over
        {d #2 d #3}}
\nc{\dx}[1]{d\,^{#1}x}
\nc{\dy}[1]{d\,^{#1}y}
\nc{\dz}[1]{d\,^{#1}z}
\nc{\dl}[1]{\frac{d\,^{#1}l}{(2\pi)^{#1}}}
\nc{\dk}[1]{\frac{d\,^{#1}k}{(2\pi)^{#1}}}
\nc{\dq}[1]{\frac{d\,^{#1}q}{(2\pi)^{#1}}}

\nc{\cc}{\mbox{$c.c.$ }}
\nc{\hc}{\mbox{$h.c.$ }}
\nc{\cf}{cf.\ }
\nc{\erfc}{{\rm erfc}}
\nc{\Tr}{{\rm Tr\,}}
\nc{\tr}{{\rm tr\,}}
\nc{\pol}{{\rm pol}}
\nc{\sign}{{\rm sign}}
\nc{\bfT}{{\bf T }}

\def\GeV{{\rm\ GeV}}

\nc{\cA}{{\cal A}}
\nc{\cB}{{\cal B}}
\nc{\cD}{{\cal D}}
\nc{\cE}{{\cal E}}
\nc{\cG}{{\cal G}}
\nc{\cH}{{\cal H}}
\nc{\cL}{{\cal L}}
\nc{\cO}{{\cal O}}
\nc{\cT}{{\cal T}}
\nc{\cN}{{\cal N}}
\nc{\rvac}[1]{|{\cal O}#1\rangle}
\nc{\lvac}[1]{\langle{\cal O}#1|}
\nc{\rvacb}[1]{|{\cal O}_\beta #1\rangle}
\nc{\lvacb}[1]{\langle{\cal O}_\beta #1 |}
\nc{\bb}{\bar{\beta}}
\nc{\bt}{\tilde{\beta}}
\nc{\ctH}{\tilde{\cal H}}
\nc{\chH}{\hat{\cal H}}
\nc{\1}{\aa}
\nc{\2}{\"{a}}
\nc{\3}{\"{o}}
\nc{\4}{\AA}
\nc{\5}{\"{A}}
\nc{\6}{\"{O}}
\nc{\al}{\alpha}
\nc{\g}{\gamma}
\nc{\Del}{\Delta}
\nc{\e}{\epsilon}
\nc{\eps}{\epsilon}
\nc{\lam}{\lambda}
\nc{\om}{\omega}
\nc{\Om}{\Omega}
\nc{\ve}{\varepsilon}
\nc{\mn}{{\mu\nu}}
\nc{\vp}{\varphi}

\nc{\advp}[3]{{\it  Adv.\ in\ Phys.\ }{{\bf #1} {(#2)} {#3}}}
\nc{\annp}[3]{{\it  Ann.\ Phys.\ (N.Y.)\ }{{\bf #1} {(#2)} {#3}}}
\nc{\apl}[3]{{\it  Appl. Phys. Lett. }{{\bf #1} {(#2)} {#3}}}
\nc{\apj}[3]{{\it  Ap.\ J.\ }{{\bf #1} {(#2)} {#3}}}
\nc{\apjl}[3]{{\it  Ap.\ J.\ Lett.\ }{{\bf #1} {(#2)} {#3}}}
\nc{\app}[3]{{\it Astropart.\ Phys.\ }{{\bf #1} {(#2)} {#3}}}
\nc{\cmp}[3]{{\it  Comm.\ Math.\ Phys.\ }{{ \bf #1} {(#2)} {#3}}}
\nc{\cqg}[3]{{\it  Class.\ Quant.\ Grav.\ }{{\bf #1} {(#2)} {#3}}}
\nc{\epl}[3]{{\it  Europhys.\ Lett.\ }{{\bf #1} {(#2)} {#3}}}
\nc{\ijmp}[3]{{\it Int.\ J.\ Mod.\ Phys.\ }{{\bf #1} {(#2)} {#3}}}
\nc{\ijtp}[3]{{\it Int.\ J.\ Theor.\ Phys.\ }{{\bf #1} {(#2)} {#3}}}
\nc{\jmp}[3]{{\it  J.\ Math.\ Phys.\ }{{ \bf #1} {(#2)} {#3}}}
\nc{\jpa}[3]{{\it  J.\ Phys.\ A\ }{{\bf #1} {(#2)} {#3}}}
\nc{\jpc}[3]{{\it  J.\ Phys.\ C\ }{{\bf #1} {(#2)} {#3}}}
\nc{\jap}[3]{{\it J.\ Appl.\ Phys.\ }{{\bf #1} {(#2)} {#3}}}
\nc{\jpsj}[3]{{\it J.\ Phys.\ Soc.\ Japan\ }{{\bf #1} {(#2)} {#3}}}
\nc{\lmp}[3]{{\it Lett.\ Math.\ Phys.\ }{{\bf #1} {(#2)} {#3}}}
\nc{\mpl}[3]{{\it  Mod.\ Phys.\ Lett.\ }{{\bf #1} {(#2)} {#3}}}
\nc{\ncim}[3]{{\it  Nuov.\ Cim.\ }{{\bf #1} {(#2)} {#3}}}
\nc{\np}[3]{{\it  Nucl.\ Phys.\ }{{\bf #1} {(#2)} {#3}}}
\nc{\npps}[3]{{\it  Nucl.\ Phys.\ Proc.\ Suppl.\ }{{\bf #1} {(#2)} {#3}}}
\nc{\pr}[3]{{\it Phys.\ Rev.\ }{{\bf #1} {(#2)} {#3}}}
\nc{\pra}[3]{{\it  Phys.\ Rev.\ A\ }{{\bf #1} {(#2)} {#3}}}
\nc{\prb}[3]{{\it  Phys.\ Rev.\ B\ }{{{\bf #1} {(#2)} {#3}}}}
\nc{\prc}[3]{{\it  Phys.\ Rev.\ C\ }{{\bf #1} {(#2)} {#3}}}
\nc{\prd}[3]{{\it  Phys.\ Rev.\ D\ }{{\bf #1} {(#2)} {#3}}}
\nc{\prl}[3]{{\it Phys.\ Rev.\ Lett.\ }{{\bf #1} {(#2)} {#3}}}
\nc{\pl}[3]{{\it  Phys.\ Lett.\ }{{\bf #1} {(#2)} {#3}}}
\nc{\prep}[3]{{\it Phys.\ Rep.\ }{{\bf #1} {(#2)} {#3}}}
\nc{\prsl}[3]{{\it Proc.\ R.\ Soc.\ London\ }{{\bf #1} {(#2)} {#3}}}
\nc{\ptp}[3]{{\it  Prog.\ Theor.\ Phys.\ }{{\bf #1} {(#2)} {#3}}}
\nc{\ptps}[3]{{\it  Prog\ Theor.\ Phys.\ suppl.\ }{{\bf #1} {(#2)} {#3}}}
\nc{\physa}[3]{{\it  Physica\ A\ }{{\bf #1} {(#2)} {#3}}}
\nc{\physb}[3]{{\it  Physica\ B\ }{{\bf #1} {(#2)} {#3}}}
\nc{\phys}[3]{{\it Physica\ }{{\bf #1} {(#2)} {#3}}}
\nc{\rmp}[3]{{\it  Rev.\ Mod.\ Phys.\ }{{\bf #1} {(#2)} {#3}}}
\nc{\rpp}[3]{{\it Rep.\ Prog.\ Phys.\ }{{\bf #1} {(#2)} {#3}}}
\nc{\sjnp}[3]{{\it Sov.\ J.\ Nucl.\ Phys.\ }{{\bf #1} {(#2)} {#3}}}
\nc{\spjetp}[3]{{\it Sov.\ Phys.\ JETP\ }{{\bf #1} {(#2)} {#3}}}
\nc{\yf}[3]{{\it Yad.\ Fiz.\ }{{\bf #1} {(#2)} {#3}}}
\nc{\zetp}[3]{{\it Zh.\ Eksp.\ Teor.\ Fiz.\  }{{\bf #1}  {(#2)} {#3}}}
\nc{\zp}[3]{{\it Z.\ Phys.\ }{{\bf #1} {(#2)} {#3}}}
\nc{\ibid}[3]{{\sl ibid.\ }{{\bf #1} {#2} {#3}}}
\nc{\rf}[1]{(\ref{#1})}
\nc{\nn}{\nonumber \\*}
\nc{\bfB}{\bf{B}}
\nc{\bfv}{\bf{v}}
\nc{\bfx}{\bf{x}}
\nc{\bfy}{\bf{y}}
\nc{\vx}{\vec{x}}
\nc{\vy}{\vec{y}}
\nc{\oB}{\overline{B}}
\nc{\oI}{\overline{I}}
\nc{\oR}{\overline{R}}
\nc{\rar}{\rightarrow}
\nc{\ti}{\times}
\nc{\slsh}{\hskip-5pt/}
\nc{\sm}{Standard~Model~}
\nc{\MP}{M_{\rm Pl}}
\nc{\tp}{t_{\rm Pl}}
\nc{\ave}{\bar{E}}

\nc{\eff}{{\rm eff}}
\nc{\kk}{\vek{k}}
\nc{\pp}{{\rm p}}
\nc{\ga}{g_{a\gamma}}
\nc{\vv}{\\}
\nc{\eee}{{\bf E}}
\nc{\bbb}{{\bf B}}
\nc{\qcd}{T_{\rm QCD}}
\nc{\G}{\rm \ G}
\def\vec#1{{\bf #1}}

\def\lae{\;^{<}_{\sim} \;} \def\gae{\; ^{>}_{\sim} \;} 

\def\ell{e^{c}LL}

\begin{document}
{\title{\vskip-2truecm{\hfill {{\small \\
	\hfill \\
	}}\vskip 1truecm}
{\LARGE  Supergravity Modification of D-term Hybrid Inflation: Solving the Cosmic String and Spectral Index Problems via a Right-Handed Sneutrino 
}}
{\author{
{\sc \large Chia-Min Lin$^{1}$ and John McDonald$^{2}$}\\
{\sl\small Cosmology and Astroparticle Physics Group, University of Lancaster,
Lancaster LA1 4YB, UK}
}
\maketitle
\begin{abstract}
\noindent

             Supergravity corrections due to the energy density of a right-handed sneutrino can generate a negative mass squared for the inflaton, flattening the inflaton potential and reducing the spectral index and inflaton energy density. For the case of D-term hybrid inflation, we show that the spectral index can be lowered from the conventional value $n = 0.98$ to a value within the range favoured by the latest WMAP analysis, $n = 0.951^{+0.015}_{-0.019}$. The modified energy density is consistent with non-observation of cosmic strings in the CMB if $n < 0.946$. The WMAP lower bound on the spectral index implies that the D-term cosmic string contribution may be very close present CMB limits, contributing at least 5\% to the CMB multipoles.  


\end{abstract} 
\vfil
 \footnoterule{\small $^1$c.lin3@lancaster.ac.uk, $^2$j.mcdonald@lancaster.ac.uk}   
 \newpage 
\setcounter{page}{1}

\section{Introduction}

          The Minimal Supersymmetric (SUSY) Standard Model (MSSM) \cite{nilles}, extended to accomodate neutrino masses, is widely regarded as the most likely candidate for a 
theory of particle physics beyond the Standard Model. Proof of the existence of the MSSM, which may be obtained by the CERN Large Hadron Collider in the near future, would strongly motivate the development of a detailed understanding of SUSY cosmology. In particular, it would be essential to develop a naturally consistent SUSY inflation model to serve as the basis for SUSY cosmology. 

           In the context of supergravity (SUGRA), SUSY inflation models based on an F-term typically suffer from the $\eta$-problem i.e. the generation of order $H^{2}$ corrections to the inflaton mass squared \cite{eta,drt1}. This problem does not arise for inflation models in which the energy density comes from a D-term\footnote{F-term inflation models with a Heisenberg symmetry \cite{oliveh}, a cancellation between SUGRA corrections and radiative corrections \cite{rmi} or a minimal K\"ahler potential with a superpotential linear in the inflaton field \cite{h2} (in particular F-term hybrid inflation models \cite{fti}) are also possible.}. In particular, D-term hybrid inflation models \cite{dti}, based on a Fayet-Iliopoulos D-term associated with a $U(1)_{FI}$ gauge symmetry, automatically evade the $\eta$-problem and so are naturally consistent with SUGRA. 
However, D-term hybrid inflation has a problem due to the formation of local cosmic strings associated with the breaking of the $U(1)_{FI}$ gauge symmetry at the end of inflation \cite{dtics}. The contribution of these cosmic strings to the 
cosmic microwave background (CMB) has been shown to be too large in models which have a naturally large value of the superpotential coupling \cite{endo}. 

          Since D-term hybrid inflation has the ability to provide a naturally consistent basis for SUSY inflation in the context of SUGRA, it is important to ask whether there is any way to overcome the cosmic string problem. One way is to reduce the superpotential coupling to less than O($10^{-4}$) \cite{endo}. However, this requires that the $U(1)_{FI}$ gauge coupling, $g$,  is much less than any known gauge coupling, $g \lae 10^{-2}$ \cite{rocher}. In this paper we will concentrate on the case of large couplings. Another approach is to reduce the mass per unit length of the cosmic strings by reducing the expectation value of the symmetry-breaking field. This requires that the energy density during inflation is lowered whilst still accounting for the observed magnitude of the primordial density perturbations. In \cite{seto} the energy density was lowered by introducing higher-order corrections to the K\"ahler potential of the D-term hybrid inflation fields. It was shown that for particular choices of the higher-order correction the slope of the inflaton potential could be flattened, thus allowing the energy density to be reduced. In \cite{kadota} an alternative approach was followed in which the D-term hybrid inflation superpotential was modified by replacing the renormalizable superpotential proportional to the inflaton with a non-renormalizable superpotential proportional to the inflaton squared. It was shown that this model could account for the observed density perturbations with a reduced energy density. An alternative approach is to prevent the formation of local cosmic strings at the end of inflation. In \cite{davis} the D-term hybrid inflation superpotential was modified by adding a second pair of $U(1)_{FI}$ charged superfields in such a way that the strings which form are semilocal. (The cosmic string problem in SUSY hybrid inflation models has recently been reviewed in \cite{jeanpost}.) 
In the context of F-term hybrid inflation, it is possible to have a non-renormalizable superpotential such that there is no symmetry-breaking phase transition, corresponding to cases of shifted and smooth hybrid inflation \cite{shift,smooth,sss}. 

          The recent three-year Wilkinson Microwave Anisotropy Probe (WMAP) results \cite{wmap3} have highlighted a second problem for the D-term hybrid inflation model, which is that the spectral index, $n$, may be too large to be consistent with CMB observations. The conventional D-term hybrid inflation model predicts that $n = 0.98$, whereas the three-year WMAP data implies that $n = 0.951^{+0.015}_{-0.019}$ at 1-$\sigma$ \cite{wmap3}. This is an even more severe problem for the modified D-term inflation model of \cite{kadota} or the small coupling limit of conventional D-term inflation, which predict that the spectral index is extremely close to 1, $n-1 \approx 10^{-5}$
\footnote{It has been suggested in \cite{aur} that conventional D-term hybrid inflation could be consistent with observation if there is a contribution to the CMB power spectrum from cosmic strings which is combined with values for the other cosmological parameters in such a way as to match to the small $n$ CMB power spectrum for the case without cosmic strings.}. 

              In this paper we present a new solution of the cosmic string and spectral index problems of D-term hybrid inflation. This solution is based on the interaction expected in the context of SUGRA between the inflaton and a massive right-handed (RH) sneutrino. We will show that for natural values of the RH sneutrino mass and expectation value it is possible for this interaction to flatten the inflaton potential sufficiently to overcome the cosmic string and spectral index problems.

             The paper is organised as follows. In Section 2 we calculate the SUGRA modification of the D-term hybrid inflation scalar potential due to a RH sneutrino. In Section 3 we solve for the evolution of the inflaton as a function of the number of e-foldings and show how the RH sneutrino modification can reduce the spectral index and energy density during inflation. Bounds on the spectral index and energy density for which the RH sneutrino-modified D-term hybrid inflation model is consistent with all CMB bounds are derived. In Section 4 we discuss the range of RH sneutrino mass and expectation value for which the cosmic string and spectral index problems are solved. In Section 5 we review the present status of the cosmic string CMB upper bound on the energy scale during inflation and estimate the contribution of cosmic strings to the CMB in the RH sneutrino-modified D-term hybrid inflation model. In Section 6 we present our conclusions.

\section{Modifying the Inflaton Potential via a RH Sneutrino}

      The most commonly studied form of neutrino mass model is that based on the see-saw mechanism \cite{seesaw}. The superpotential responsible for the light neutrino masses is
\be{ne1} W_{\nu} = \lambda_{\nu} \Phi H_{u} L  + \frac{M_{\Phi}}{2} \Phi^{2}   ~,\ee
where $\Phi$ is the RH neutrino superfield, $H_{u}$ and $L$ are the Higgs and lepton superfield doublets and $M_{\Phi}$ is the RH neutrino mass. In this we have suppressed generation indices. In the following we will consider the case where only a single RH sneutrino is given a large expectation value during inflation.   
 
       The superpotential of D-term hybrid inflation is given by \cite{dti} 
\be{e2}   W_{D} = \lambda S \Phi_{+} \Phi_{-}         ~,\ee
where $S$ is the inflaton superfield and $\Phi_{\pm}$ are superfields charged under the $U(1)_{FI}$ gauge symmetry responsible for the Fayet-Iliopoulos term.   
The corresponding SUSY scalar potential is
\be{e3}  V(S, \Phi_{+}, \Phi_{-}) = \lambda^{2} \left[ \left|S\right|^{2}\left(\left|\Phi_{+}\right|^{2} 
+ \left|\Phi_{-}\right|^{2} \right) + \left|\Phi_{+}\right|^{2}\left|\Phi_{-}\right|^{2} \right] 
+ \frac{g^{2}}{2} \left( \left|\Phi_{+}\right|^{2} - \left|\Phi_{-}\right|^{2} + \xi \right)^{2} 
   ~,\ee
where $\xi$ is the Fayet-Iliopoulos term. 
$\Phi_{+}$ and $\Phi_{-}$ are equal to zero at the minimum of the potential as a function of $|S|$ when $|S| \gg |S|_{c} = g \xi^{1/2}/\lambda$. The 1-loop inflaton potential is then \cite{rocher} 
\be{e4}  V(S) = V_{o} + \frac{g^{4} \xi^{2}}{16 \pi^{2}} \left[ \ln \left( \frac{|S|^{2}}{\Lambda^{2}} \right) 
+ \left(z + 1\right)^{2} \ln \left(1 + z^{-1} \right) +  \left(z - 1\right)^{2} \ln \left(1 - z^{-1} \right)
\right]    ~,\ee 
where $V_{o} = g^{2} \xi^{2}/2$ and $z = \lambda^{2} |S|^{2}/g^{2} \xi^{2}$. The effect of SUGRA is to replace 
$|S|^{2}$ by $|S|^{2}e^{|S|^{2}/M^{2}}$ \cite{rocher}, where $M = M_{Pl}/\sqrt{8 \pi}$. In this paper we will restrict attention to values $g \approx 0.1$ and $\lambda \gae 0.1$, in which case $|S|^{2} \ll M^{2}$ and $z \gg 1$.  ($|S|^{2}/M^{2}  =  g^{2}N/4\pi^{2}$ in the standard D-term hybrid inflation model at $N$ e-foldings before the end of inflation \cite{dti}.) In this case the 1-loop potential can be approximated by
 \be{e4}  V(S) = V_{o} +  \frac{g^{4} \xi^{2}}{16 \pi^{2}} \ln \left( \frac{|S|^{2}}{\Lambda^{2}} \right)   ~.\ee

      In the following we will consider the case where there are no RH neutrino superpotential terms of the form $\Phi^{r}$ with $r \neq 2$. Such terms can be eliminated by an R-symmetry, with R-charges as given in Table 1. 
 \begin{table}[h]
 \begin{center}
 \begin{tabular}{|c|c|c|c|c|c|c|c|}
	\hline   $u^{c}$ &  $d^{c}$ &  $e^{c}$ & $Q$ & $L$ & $H_{u}$ & $H_{d}$ & $\Phi$  \\ 
	\hline   1 & 1 & 1 & 1 & 1  & 0 & 0 & 1\\
	\hline     
 \end{tabular}
 \caption{\footnotesize{R-charge assignment.}}  
 \end{center}
 \end{table}
In this case the SUSY scalar potential of the RH sneutrino is entirely due to its mass term, 
\be{ne4}  V(\Phi) = M_{\Phi}^{2} |\Phi|^{2}     ~.\ee
The R-symmetry is broken to R-parity by soft SUSY breaking gaugino masses and scalar interactions characterized by a mass scale $m_{s}$ and by the $\mu H_{u}H_{d}$ term\footnote{Other R-charge assignments are possible, such as $R(u^{c},d^{c},e^{c},H_{u},H_{d},\Phi) = 1$, $R(Q,L) = 0$, under which the superpotential including the $\mu H_{u}H_{d}$ term is completely R-symmetric. The R-charge assignment in Table 1 corresponds to the case where R-parity is a discrete unbroken subgroup of the R-symmetry.} 
in the MSSM superpotential 
\cite{nilles}. However, no significant new RH neutrino superpotential terms will be introduced so long as all R-symmetry breaking terms are proportional to powers of $\mu \approx m_{s} \approx 100 \GeV \ll M_{\Phi}$. The assumption of R-symmetry suppression of superpotential terms is consistent with the assumption made in D-term hybrid inflation that there are no non-renormalizable corrections to the inflaton superpotential of the form $S^{r}$, which may be understood as due to an R-symmetry under which only superpotential terms linear in $S$ are permitted.

          An F-term contribution to the scalar potential in SUGRA generally results in mass squared terms being induced for all scalars \cite{eta,drt1}. Therefore we expect that a RH sneutrino will induce significant SUGRA corrections in the inflaton potential if it has a large enough expectation value. We will consider a K\"ahler potential for the inflaton and RH sneutrino of the form 
\be{e5}  K = S^{\dagger}S + \Phi^{\dagger}\Phi + \frac{c S^{\dagger}S\Phi^{\dagger}\Phi}{M^{2}}       ~.\ee 
In this we have assumed that the natural mass scale of higher-order corrections to the K\"ahler potential is $M$, so that 
$|c|$ is O(1), and we have included only those higher-order corrections which play an important role in modifying the inflaton potential \footnote{Higher order K\"ahler corrections involving only the inflaton field, such as $\gamma (S^{\dagger}S)^{2}/M^{2}$ with $|\gamma| \approx 1$, will not signficantly modify the inflaton potential so long as $|S|^{2}/M^{2}$ is small compared with 1. The full effect of higher-order K\"ahler corrections will be considered in future work.}. The SUGRA scalar potential is then given by 
\be{e6} V = e^{\frac{K}{M^{2}}}\left[  \left(W_{m} + \frac{W K_{m}}{M^{2}} \right)^{\dagger} K^{m^{\dagger}n} \left(W_{n} + \frac{W K_{n}}{M^{2}} \right)    - \frac{3 \left|W\right|^{2}}{M^{2}}    \right]
~,\ee
where $W = W_{D} + W_{\nu}$ and $K^{m^{\dagger}n}$ is the inverse matrix of 
$$K_{m^{\dagger}n} = 
\partial^{2}K/\partial \phi_{m}^{\dagger} \partial \phi_{n}  ~.$$  
Assuming that $|S|$, $|\Phi| < M$ and expanding in $|S|/M$ and $|\Phi|/M$, the leading order correction to the inflaton potential due to 
the RH sneutrino from \eq{e6} is given by 
\be{e7}   \Delta V  =  
- \frac{ \left(c-1\right) M_{\Phi}^{2} |\Phi|^{2} |S|^{2}}{M^{2}}  ~.\ee 
The inflaton mass squared correction is negative when $c > 1$. In this case it is possible for the slope of the inflaton potential to be flattened, reducing both the energy density during inflation and the spectral index\footnote{Similar corrections have recently been considered in the context of F-term hybrid inflation models 
\cite{shaf2}.}.  

            Assuming that the inflaton expectation value is real, $S \equiv Re(S) = s/\sqrt{2}$, the inflaton potential including RH sneutrino corrections is therefore given by 
\be{e10} V(s) = V_{o} + \frac{\alpha}{2} \ln \left( \frac{s^{2}}{2 \Lambda^{2}} \right)  - \frac{\kappa s^{2}}{2}    ~,\ee
where we have defined $\alpha$ and $\kappa$ by  
$$    \alpha = \frac{g^{4} \xi^{2}}{8 \pi^{2}} \;\;,\;\;\;\; \kappa = \frac{ \left(c-1\right) M_{\Phi}^{2} |\Phi|^{2} }{M^{2}}           ~.$$
The possibility that D-term hybrid inflation could be modified by a mass squared term has been previously discussed in \cite{lrdterm} (for the case of a positive mass squared term) and in \cite{lbdterm} (for the case of a negative mass squared term, motivated by the possibility of hilltop inflation). Although these terms were added phenomenologically, their possible origin from an F-term in the scalar potential was noted. Mass squared terms have also been considered in the context of F-term hybrid inflation models \cite{ss1,yamayoko,shaf1,shaf2}.

\section{D-term Inflation with RH Sneutrino Modification}

         In this section we solve for the evolution of $s$ as a function of the number of e-foldings until the end of inflation, $N$, and calculate the modification of the spectral index and energy density.   

          In the slow-roll approximation the $s$ field equation becomes 
\be{e11} 3 H \dot{s} = - \frac{\partial V}{\partial s} \equiv - \left( \frac{\alpha}{s} - \kappa s \right)        ~.\ee
In the following we will consider the case where $\kappa$ remains approximately constant during inflation. In this case it is possible to obtain analytic solutions for the $s$ evolution and associated cosmological parameters. Using $\partial s/\partial t = - H \partial s /\partial N$,  \eq{e11} can be written as 
\be{e11a} 3 H^{2} \frac{\partial s}{\partial N} =  \frac{\kappa}{s} \left(s_{o}^{2} - s^{2} \right)  \;\; ; \;\;\; 
s_{o}^{2} = \frac{\alpha}{\kappa}  ~.\ee 
Integrating this gives the $s$ evolution as a function of $N$,
\be{e12}   s^{2}(t) = s^{2}_{o} \left(1 - e^{-\gamma} 
\right) \;\;;\;\;\;\;   \gamma = 2 \kappa N/3 H^{2}   ~,\ee
where we have assumed that $H \approx (V_{o}/3 M^{2})^{1/2}$ is constant.   
In the limit where $ \gamma \ll 1$ this reduces to the standard 
D-term inflation expression, $s^{2}/M^{2} = g^{2}N/2 \pi^{2}$, whilst for $\gamma \gg 1$,
$s_{N}^{2} \approx s_{o}^{2}$. We will see that the latter case is ruled out by its
large deviation from scale-invariance.

\subsection{Solution of the D-term Inflation Spectral Index Problem}

         The spectral index is given by the standard expression \cite{lr}, 
\be{e13}     n = 1 + 2 \eta - 6  \epsilon             ~\ee 
where $\eta = M^{2} V^{''}/V$ and $\epsilon = (M^{2}/2)(V^{'}/V)^{2} $. 
Using the above potential and slow-roll solution we obtain 
\be{e13a}   \eta = -\frac{\gamma}{2 N} \left(\frac{1}{\left(1 - e^{-\gamma}\right)} + 1\right) \; ; \;\;\; 
\epsilon =  \frac{g^{2}}{16 \pi^{2} N} \frac{\gamma e^{-2 \gamma}}{\left(1 - e^{-\gamma} \right)}           ~.\ee
The contribution of $\epsilon$ to $n$ is negligible compared with that of $\eta$. Therefore
\be{e14} n = 1 - \frac{\gamma}{N} \left(\frac{1}{\left(1 - e^{-\gamma}\right)} + 1 \right)  ~.\ee 
In the small $\kappa$ limit ($ \gamma \ll 1$) this reduces to the standard 
D-term inflation result, $n = 1 - 1/N$. The effect of the RH sneutrino correction is to increase the deviation of $n$ from 1. In Figure 1 we show the value of $n$ as a function of $\gamma$ when $N = 60$. We also show the recent three-year 1-$\sigma$ WMAP limits on $n$, which imply that $n$ lies in the range 0.932 to 0.966, with a best-fit value of 0.951.
This strongly disfavours conventional D-term inflation, for which $n = 0.983$, but allows RH sneutrino-modified D-term inflation if $0.66 < \gamma < 1.87$.  At values of $\gamma \gg 1$ the spectral index is given by $ n \approx 1 - 2\gamma/N$. Therefore the spectral index becomes highly scale-dependent as $\gamma$ increases to values large compared with 1.

\begin{figure}[h] 
                    \centering                   
                    \includegraphics[width=0.50\textwidth, angle=-90]{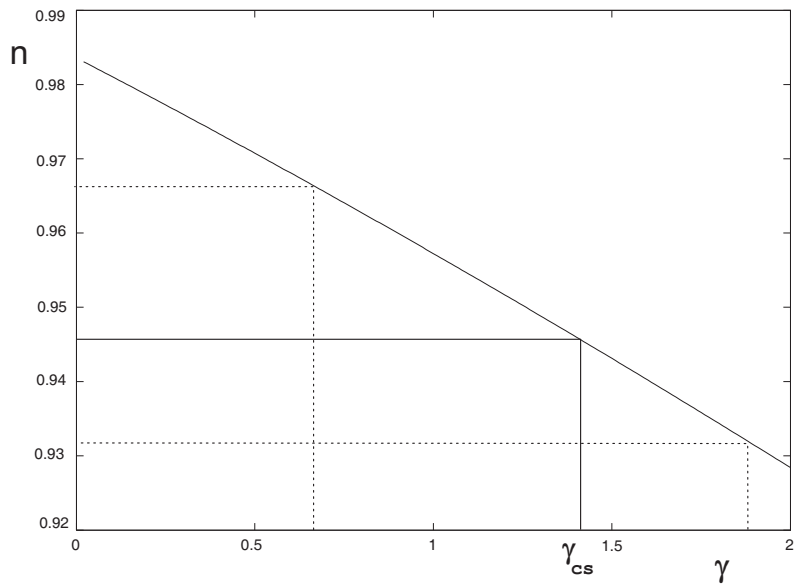}
                    \caption{\footnotesize{Spectral index as a function of $\gamma$. The WMAP upper and lower bounds on 
                            $n$ and $\gamma$ are shown as broken lines. The solid line is the cosmic string lower bound on $\gamma$, $\gamma_{cs}$.}}
                    \end{figure}

\subsection{Curvature Perturbation and Solution of the D-term Inflation Cosmic String Problem} 

         The curvature perturbation is given by \cite{lr} 
\be{e15}    \frac{4}{25} P_{\zeta} = \frac{1}{75 \pi^{2} M^{6}} \frac{V^{3}}{V^{'\;2}}                ~.\ee  
Using the potential, \eq{e10}, and the slow-roll solution for $s$, \eq{e12}, we obtain 
\be{e17} P_{\zeta} =  \frac{N \xi^{2}}{3 M^{4}} \frac{1}{\Gamma}    ~,\ee 
where 
\be{e18} \Gamma  = \frac{\gamma e^{-2 \gamma} } { \left( 1  - e^{-\gamma} \right)}   ~.\ee  
In the limit $\gamma \ll 1$, $P_{\zeta}$ reduces to the conventional D-term inflation result, $P_{\zeta} = N \xi^{2}/3 M^{4}$. In this case 
the observed curvature perturbation, $P_{\zeta}^{1/2} = 4.8 \times 10^{-5}$, is obtained when 
\be{e19}    \xi^{1/2} = 7.9 \times 10^{15} \left(\frac{60}{N}\right)^{1/4} \GeV      ~.\ee
This is incompatible with the upper limit from the non-observation of cosmic strings in the CMB, 
$\xi^{1/2} < 4.6 \times 10^{15} \GeV$ \cite{endo}. 
The RH sneutrino correction modifies the value of $\xi^{1/2}$ needed to explain the curvature perturbation to 
\be{e20}    \xi^{1/2} = 7.9 \times 10^{15} \Gamma^{1/4} \left(\frac{60}{N}\right)^{1/4} \GeV      ~.\ee
Compatibility with the cosmic string upper bound therefore requires that $\Gamma < 0.12$.  
This is achieved if $\gamma > \gamma_{CS} = 1.4$  

           As seen from Figure 1, requiring that $\gamma > 1.4$ implies that there is an upper bound on the spectral index, $n < 0.946$. Therefore both the spectral index problem and cosmic string problem of D-term inflation can be solved if $1.4 < \gamma < 1.87$, in which case $0.932 < n < 0.946$. In Figure 2 we plot the value of $\xi^{1/2}$ as a function of $n$. From this we see that the WMAP and cosmic string constraints imply that $3.8 \times 10^{15} \GeV < \xi^{1/2} < 4.6 \times 10^{15} \GeV$.

\begin{figure}[h] 
                    \centering                   
                    \includegraphics[width=0.50\textwidth, angle=-90]{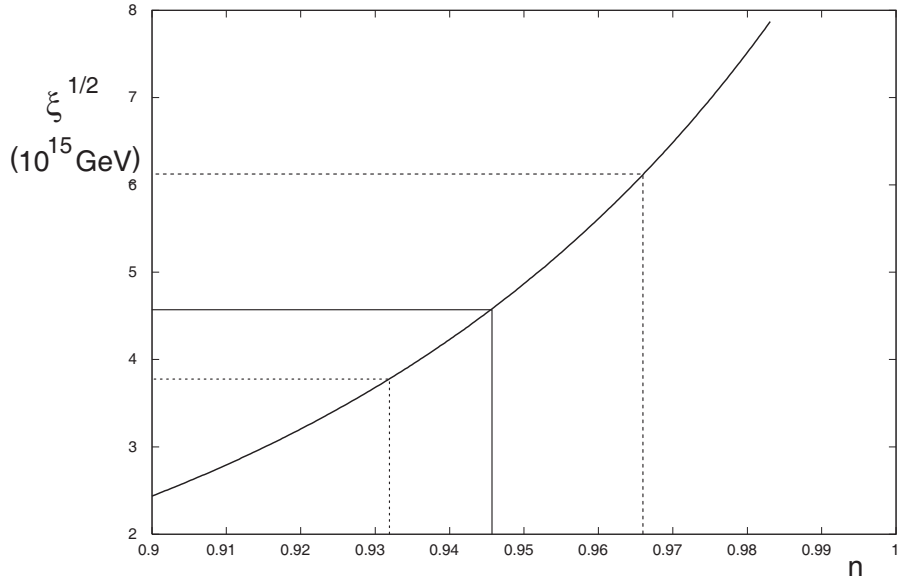}
                    \caption{\footnotesize{$\xi^{1/2}$ as a function of $n$. The upper and lower bounds on $n$ from WMAP are shown as broken lines. The solid line is from the cosmic string upper bound on $\xi^{1/2}$.}}
                    \end{figure}

\section{RH Sneutrino Properties and the Modification of the D-term Potential}

                    In the previous section we calculated the range of $\gamma$ for which it is possible to have a successful D-term hybrid inflation model. It was assumed that $\kappa$ is approximately constant when length scales relevant to cosmology are exiting the horizon, which in turn requires that the RH sneutrino is slow-rolling during this period. This is also necessary in order to have a spectral index which is not varying rapidly with scale. It was also assumed that $|\Phi| < M$ so that we can expand the SUGRA scalar potential as a power series in $|\Phi|/M$.

     $\gamma \geq \gamma_{CS} = 1.4$ is necessary in order to successfully modify the D-term hybrid inflation model.  $\gamma = \gamma_{CS}$ is satisfied when  
\be{ne10}    M_{\Phi}^{2} = \frac{ 3 \gamma_{CS} M^{2} H^{2}}{2 N \left(c-1\right) |\Phi|^{2}}         ~.\ee
The condition for the RH sneutrino to be slow-rolling, $V^{''}(\Phi) \lae H^{2}$, is satisfied if  $M_{\Phi} \lae H$. 
Therefore $\gamma = \gamma_{CS}$ and $M_{\Phi} \lae H$ are simultaneously satisfied if   
\be{ne11}   |\Phi| \gae  \left( \frac{ 3 \gamma_{CS}}{2 N \left(c-1\right)} \right)^{1/2} M       ~\ee
With $N = 60$ this requires that $|\Phi| \gae 0.19 M/\left(c-1\right)^{1/2}$.  
Therefore the D-term hybrid inflation model may be 
successfully modified by a slow-rolling RH sneutrino field when $|\Phi|$ is in the range (0.1-1)$M$. For each value of $|\Phi|$ in this range there is a corresponding value of $M_{\Phi}$ given by \eq{ne10} for which D-term hybrid inflation model is successfully modified.   

                 The required range of values of $|\Phi|$ is of the order of $M$, which is a natural mass scale in SUGRA models. Is there
any way to understand why the value of $|\Phi|$ at $N \approx 60$ only slightly modifies the inflaton potential?  
One possible way to understand this is to note that, since $M_{\Phi} \lae H$, the RH sneutrino potential during inflation is effectively featureless up to the value at which SUGRA corrections become important, $|\Phi| \approx M$. In this case it is possible that,
in a given horizon-sized domain at $N \approx 60$, all values of $|\Phi|$ up to $|\Phi| \approx M$ are equally probable. The most likely value for $|\Phi|$ in a domain chosen at random would then be in the range (0.1-1)$M$, in which case the value of $|\Phi|$ could just happen to slightly modify the inflaton potential.  A related possibility is to note that once $|\Phi|$ is larger than the value $|\Phi|_{CS}$ for which $\gamma = \gamma_{CS} = 1.4$, the deviation from scale-invariance rapidly increases with increasing $|\Phi|$, since $\gamma \propto |\Phi|^{2}$. (For example, $n = 0$ once $|\Phi| = 4.6 |\Phi|_{CS}$.) Therefore a highly scale-dependent red spectrum  of primordial density perturbations ($n \ll 1$) will occur once $|\Phi|$ is significantly larger than $|\Phi|_{CS}$. This would result in a completely different structure formation scenario as compared with that of the observed Universe. This altered structure formation scenario could serve as an anthropic cut-off, implying that the most probable value of $|\Phi|$ in a given horizon-sized domain at $N \approx 60$ is close to the value at which the inflaton potential and spectral index are only slightly modified \footnote{This is similar to the argument given in \cite{lindeax} for the small value of the axion field in theories with a large axion decay constant.}. 

      From \eq{ne10} it follows that the value of $M_{\Phi}$ must be larger than $0.19 H/\left(c-1\right)^{1/2}$ when $|\Phi| < M$. Since $H = 3.6 \times 10^{11} \GeV$ in the D-term inflation model with $g = 0.1$ and $\xi^{1/2} = 4.6 \times 10^{15} \GeV$, the RH neutrino mass associated with the modification of the inflaton potential must typically be of the order of $10^{11} \GeV$. It is interesting to note that this is a typical RH neutrino mass scale in see-saw models of neutrino mass \cite{nmass}.  

          So far we have considered the density perturbations to be due to quanum fluctuations of the inflaton field. However, since $M_{\Phi} < H$, quantum fluctuations of the RH sneutrino field may also contribute to the density perturbations. So long as the RH sneutrino decays to radiation before the inflaton coherent oscillations decay, there will be no isocurvature component. However, the adiabatic perturbation may be influenced by the RH sneutrino field. We will return this issue in future work, but here we will argue that the effect of the RH sneutrino fluctuations is typically neglibible. The only existing analysis of the effect of a second scalar field in D-term hybrid inflation is given in \cite{kt}, which considered the effect of fluctuations a generic flat direction field on the adiabatic density perturbations. It was shown that so long as $|\dot{s}| \leq |\dot{\phi}|$ and the potential is dominated by $V_{o}$, the effect of the flat direction scalar on the adiabatic density perturbation will be negligible. For slow-rolling fields this condition becomes $|V^{\;'}(s)| < |V^{\;'}(\phi)|$. In our case
\be{qf1}  V^{\;'}(s) = \frac{\alpha}{s} - \kappa s \equiv \frac{\omega \alpha}{s} ~,\ee
where $\omega < 1$ since the effect of the RH sneutrino correction is to flatten the inflaton potential, and 
\be{qf2}  V^{\;'}(\phi) = M_{\Phi}^{2} \phi   ~.\ee
From \eq{ne10} and \eq{ne11} it follows that $M_{\Phi}^{2} \gae 0.1 H^{2}$ and $\Phi \gae 0.1 M$, so $V^{\;'}(\phi) \gae 0.01 H^{2}M$. The solution for $s(t)$, \eq{e12}, implies that 
\be{qf3} \frac{\omega \alpha}{s} = \frac{3 g \omega}{2 \pi \sqrt{2N}} 
\left( \frac{\gamma}{1- e^{-\gamma}} \right)^{1/2} H^{2} M     ~.\ee
With $\gamma = \gamma_{CS} = 1.4$ and $N = 60$ this gives $V^{\;'}(s) = 0.06 g \omega H^{2} M$. So with $g \lae 0.1$ amd $\omega < 1$ we expect that $V^{\;'}(s) \lae 5 \times 10^{-3} H^{2} M$. Therefore the adiabatic density perturbation will typically be dominated by the inflaton fluctuations.

\section{Cosmic String Bounds on $\xi^{1/2}$ in D-term Hybrid Inflation}

       In the above we have used the cosmic string upper bound $\xi^{1/2} < 4.6 \times 10^{15} \GeV$,  
as estimated in \cite{endo}. This bound comes from the observational CMB requirement that the cosmic string contribution to the multipole moments, $C_{l}$, is less than approximately $10 \%$ of the total. 
The main source of uncertainty in the cosmic string upper bound is the relation between the cosmic string mass per unit length and the contribution to the CMB, which can only be established via high-resolution numerical simulations. The 
above upper bound is based on the results of \cite{allen}, which found a $l$-independent ('flat') contribution to 
CMB multipoles, $C_{l}$, corresponding to a COBE-normalised value (the value at which all the CMB power is due to cosmic strings) $G\mu \approx 1.7 \times 10^{-6}$. Here $\mu$ is the string energy per unit length, given in D-term hybrid inflation by $\mu = 2 \pi \xi$. A more recent simulation \cite{shellard} has found a lower value of $G \mu$ , $G\mu = (0.7 \pm 0.2) \times 10^{-6}$, whilst in \cite{wyman}, recently updated in \cite{wyman2}, it was estimated that $G \mu \approx 1.1 \times 10 ^{-6}$. Because of the narrow range of values of $\xi^{1/2}$ which can simultaneously suppress the cosmic string contribution and give a spectral index within the latest WMAP limits, $3.8 \times 10^{15} \GeV < \xi^{1/2} < 4.6 \times 10^{15} \GeV$, a definitive reduction of the CMB upper bound by more than a factor 
of 0.7 would rule out the RH sneutrino-modified D-term hybrid inflation model. However, given the difficulty in accurately simulating the cosmic string contribution to the CMB, the range of quoted results and the need to specifically simulate for the case of D-term hybrid inflation, it seems reasonable to conclude that the allowed range of $\xi^{1/2}$ values has not yet been excluded. 

                The spectral index lower bound on $\xi^{1/2}$ suggests that the cosmic string contribution to the CMB must be very close to the present observational limits. If we assume that the upper limit used in this paper is a accurate estimate of the value of $\xi^{1/2}$ at which the string contribution, $C_{l}^{str}$, is approximately $10 \%$ of $C_{l}$, then since $C_{l}^{str}$ is proportional to  $\xi^{2}$ \cite{endo}, the lower bound on $\xi^{1/2}$ from the spectral index implies a lower limit on the cosmic string contribution to $C_{l}$ given by $10\% \times (3.8/4.6)^{4} \approx 5 \%$.

\section{Conclusions}

              We have shown that SUGRA corrections due to a RH sneutrino can modify the D-term hybrid inflation scalar potential in such a way that the energy density during inflation is consistent with non-observation of cosmic strings in the CMB and the spectral index is within the limits favoured by three-year WMAP data. In order to successfully modify the inflaton potential, the RH sneutrino field must be the range (0.1-1)$M$ and the associated RH neutrino mass must typically be of the order of $10^{11} \GeV$.   

                 Our results suggest that D-term hybrid inflation cosmic strings should contribute at least 5$\%$ of the CMB multipoles. The narrow window of values of inflaton energy scale $\xi^{1/2}$ determined by the cosmic string upper bound and the spectral index lower bound emphasize the need for a definitive upper bound on the string mass per unit length from high-resolution simulations of the cosmic string contribution to the CMB.

        The analysis presented here is consistent for gauge coupling $g \lae 0.1$ and RH sneutrino mass $M_{\Phi}$ significantly smaller than $H$. In this case we can analytically solve for the evolution of the inflaton and obtain expressions for the spectral index and the primordial density perturbation. It is possible that larger $g$ and $M_{\Phi}$ can successfully modify the inflaton potential, but this would require a numerical solution of the field equations taking into account SUGRA corrections to the inflaton potential and the time-dependence of the RH sneutrino field. In this case there is also the possibility of a running spectral index due to the evolution of the RH sneutrino field during inflation. RH sneutrino corrections may also play a role in other inflation models which suffer from a spectral index problem, such as the non-renormalizable D-term hybrid inflation model of \cite{kadota}. We will consider these issues in future work \cite{lin2}.

\end{document}